# Variable optical chirality in an atomic assisted microresonator

**Hao Zhang,[2] Wenxiu Li,[1] Xiaoyang Chang,[1] Peng Han,[1] Shuo Jiang[1], Yang Zhou[2], Anping Huang,[1] Zhisong Xiao[1,3]\***

[1]*Key Laboratory of Micro-nano Measurement, Manipulation and Physics (Ministry of Education), School of Physics, Beihang University, Beijing 100191, China*
[2]*Frontier Research Institute of Innovative Science and Technology, Beihang University, Beijing 100191, China*
[3]*Beijing Academy of Quantum Information Sciences, Beijing 100193, China*
*\*Corresponding author: zsxiao@buaa.edu.cn*

A whispering gallery mode resonator with a cavity-made slot filled atomic vapor is demonstrated, in which the chiral symmetry is broken induced by asymmetric backscattering of counter-propagating optical waves in the WGM microcavity. The chirality of resonator modes is steered via tuning dispersion relation in the cavity-made slot. The displacement sensitivity of proposed system in response to the length variation of cavity-made slot is also analysed. The system exhibiting a high displacement sensitivity 15.22 THz/nm could be applied to optomechanical sensing applications.

## 1. INTRODUCTION

Optical micro-resonators with high-Q factors and small mode volumes for sensing, which can provide significant enhancement for light-matter interactions, have received considerable attraction over the last decade years, including temperature [1-5], refractive index [6, 7], pressure [8], biosensing [9-11] and rotation sensors [12-18]. Traditionally, there are two measurement methods to detect perturbations for an optical whispering gallery mode (WGM) microcavity. One is that the measurement point is fixed on the maximum slope value of the transmission or reflection spectrum, when the system is perturbed by a detectable change there will be a maximum output intensity change. The other measurement method is monitoring the variation of resonant frequency induced by a perturbation. Besides that, although mode-shift measurement is of a large sensing range, mode broadening detection presents a self-reference sensing signal which is immune to various noises including environmental temperature fluctuations and frequency drift of probe laser [19].

Recently, most of studies focus on non-Hermitian systems with loss and gain, which exhibit special spectral degeneracies, exceptional point (EP). At an EP two or more eigenstates and corresponding eigenstates coalesce [20]. In the vicinity of an EP, the clockwise (CW) and counterclockwise (CCW) travelling modes exhibit strong chiral behavior asymmetry with an unbalanced contribution, due to the asymmetric backscattering of counter-propagating optical waves. In an open quantum system, deformed microdisks are first utilized to realize chiral behaviour and directional light emission [21-24]. In addition to controlling the flow of light and laser emission, Yang used two nano-tips as Rayleigh scatterers to steer optical cavities at EPs for enhanced nanoparticle sensing via observing the mode splitting in the transmission spectrum [25-28]. Operating chiral modes in WGM microcavities have important implications to enable high-performance on-chip sensors.

In this paper, we demonstrate in a WGM resonator with a cavity-made slot filled atomic vapor the chiral modes are manipulated via dispersion variation in the cavity-made slot.

## 2. THEORETICAL MODE AND ANALYSIS

The proposed schematic is that a whispering gallery mode (WGM) resonator with an intrinsic loss rate $\kappa_0$ couples the evanescent light with a coupling rate $\kappa_{ex}$ through an input and output taper fiber, as depicted in Fig.1(a). A cavity-made slot fills an atomic vapor along the radial direction on the edge of the toroidal microresonator. The cavity-made slot is perpendicular to the light propagating inside the ring resonator, therefore, the dominant effect of this slot is cross-coupling rate $\gamma_{c1,2}$ of clockwise and counterclockwise propagating WGMs, where subscript 1 and 2 represent CW cross-coupling into CCW and CCW into CW, respectively.

The Heisenberg equations of motion describing the evolution of the propagating CW and CCW cavity modes [29],

$$\dot{a}_{cw} = -[i\Delta\omega + i\gamma_{c1} + \gamma_{total}]a_{cw} - [i\gamma_{c1} + \frac{\Gamma_{slot}}{2}]a_{ccw} - \sqrt{\kappa_{ex}}a_{in}$$

$$\dot{a}_{ccw} = -[i\Delta\omega + i\gamma_{c2} + \gamma_{total}]a_{ccw} - [i\gamma_{c2} + \frac{\Gamma_{slot}}{2}]a_{cw},$$

(1)

$$T = \left|\frac{a_{cw}^{out}}{a_{in}}\right|^2 = \left|1 + \frac{\kappa_{ex}(i\Delta\omega + i\gamma_{c2} + \gamma_{total})}{(i\gamma_{c1} + \Gamma_{slot}/2)(i\gamma_{c2} + \Gamma_{slot}/2) - (i\Delta\omega + i\gamma_{c1} + \gamma_{total})(i\Delta\omega + i\gamma_{c2} + \gamma_{total})}\right|^2$$

$$R = \left|\frac{a_{ccw}^{out}}{a_{in}}\right|^2 = \left|\frac{-\kappa_{ex}(i\gamma_{c2} + \Gamma_{slot}/2)}{(i\gamma_{c1} + \Gamma_{slot}/2)(i\gamma_{c2} + \Gamma_{slot}/2) - (i\Delta\omega + i\gamma_{c1} + \gamma_{total})(i\Delta\omega + i\gamma_{c2} + \gamma_{total})}\right|^2.$$

(2)

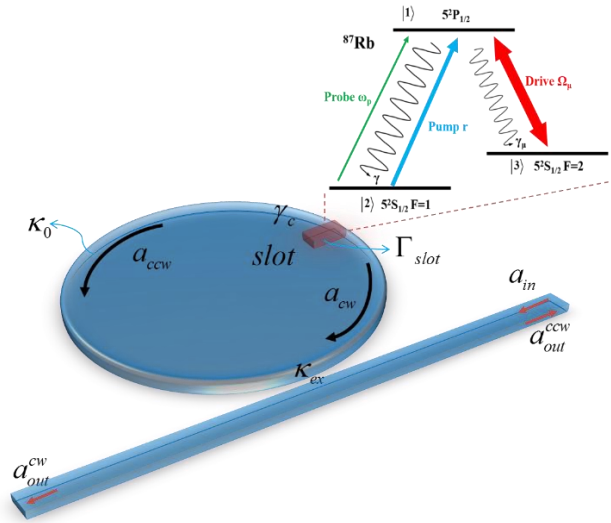

Fig. 1. (a) The schematic diagram of a WGM cavity coupled a cavity-made slot by filling a three-level atomic vapor.

For the cavity-made slotted resonator, the cross-coupling rate $\gamma_{c1,2}$ is relevant to the slot effective field reflectivity $R^{1,2}_{slot}$ and the total loss rate $\gamma_{total}$ of the resonator coupled system, and $\gamma_{c1,2}$ could be expressed by

$$\gamma_{c1,2} = \frac{|R^{1,2}_{slot}|}{\tau_0} - \gamma_{total},$$

(3)

where $\tau_0 = 2\pi n_{ring}R_0/c$ is the optical round-trip time in the resonator with radius $R_0$. The propagating WGM (CW or CCW) is reflected by the cavity-made slot, which induces cross-coupling between clockwise and counterclockwise beams. Therefore, due to the reflection of the cavity-made slot, the degenerate CW and CCW modes couple to each other and generate two new eigenmodes with eigenfrequencies

$$\omega_{\pm} = \Delta\omega + (\frac{\gamma_{c1} + \gamma_{c2}}{2}) - i\gamma_{total}$$
$$\pm\sqrt{\frac{1}{4}(\gamma_{c1} - \gamma_{c2})^2 + (\gamma_{c1} - i\Gamma_{slot}/2)(\gamma_{c2} - i\Gamma_{slot}/2)}.$$

(4)

where $\Delta\omega = \omega - \omega_c$ is the optical frequency detuning from the resonant angular frequency $\omega_c$. $\gamma_{total} = (\Gamma_{slot} + \kappa_0 + \kappa_{ex})/2$ describes the total loss rate of the coupled system, and $\Gamma_{slot}$ the absorption loss rate derived from atom vapor in the cavity-made slot. Owing to cavity-made slot, a partial light reflects on the slot surface, and a pair of degenerate modes supported in WGM cavity turn to two non-degenerate WGMs. According to the input-output relation through taper fiber $a_{cw}^{out} = a_{in} + \sqrt{\kappa_{ex}}a_{cw}$, $a_{ccw}^{out} = \sqrt{\kappa_{ex}}a_{ccw}$, the transmission and reflection spectra are obtained as

The dispersion relation has been discussed in [30]. For dilute gaseous medium $|\chi| \ll 1$, the relations between refractive index and susceptibility are

$$\text{Re}(n_{slot}) \approx 1 + \frac{1}{2}\text{Re}(\chi)$$
$$\text{Im}(n_{slot}) = \frac{\pi c \alpha_{slot}}{\omega_p} \approx -\frac{1}{2}\text{Im}(\chi),$$

(5)

where $\alpha_{slot}$ is loss coefficient and $a_{slot} = \exp(-\alpha_{slot}d/2)$ is propagation amplitude attenuation. The absorption of atomic vapor into the cavity-made slot induces the slot's loss rate $\Gamma_{slot}$, thus $\Gamma_{slot}$ can be expressed by

$$\Gamma_{slot} = \frac{c\alpha_{slot}}{2n_{slot}} = -\frac{\omega_p \text{Im}(\chi)}{4\pi n_{slot}}.$$

(6)

The reflectivity of the cavity-made slot is given by

$$R^{1,2}_{slot} = \left|r^{1,2}_{mirr} + \frac{T^{1,2}_{mirr}r^{2,1}_{mirr}a \cdot e^{2i(d\omega_{12}/c)[1-(\Delta/\omega_{12})]n_{slot}(\Delta)}}{1 - r^1_{mirr}r^2_{mirr}a \cdot e^{2i(d\omega_{12}/c)[1-(\Delta/\omega_{12})]n_{slot}(\Delta)}}\right|^2.$$

(7)

The superscript 1 and 2 represent the reflectivity on the both side of the cavity-made slot. The probe optical field with the wavelength 795nm is designed to satisfy the resonance condition in the WGMs toroidal microresonator. The reflectivity of the slot surface $R^2_{mirr}$ is set to 0.999 and $\lambda_0$=795nm, $\gamma$=9.2π×10$^6$, $AA=\gamma$ throughout this paper unless specified, and $r^{1,2}_{mirr} = \sqrt{R^{1,2}_{mirr}}$.

Fig.2 delineates the slot reflection to CCW and CW waves vary as a function of the reflection of slot surface $R^1_{mirr}$ with different atomic pump rate $r$ and driving field Rabi frequency $\Omega_\mu$. As discussed in [29], provided $r$ is larger than $\gamma$, anomalous dispersion is generated between two gain peaks which is analogous to the case of a double Raman pumped gain medium. Conversely, if $r$ is less than $\gamma$ normal dispersion is generated between two absorption peaks. The reflection rate $R^2_{slot}$ of CW wave declines slowly from one and has an obviously decrease when $R^1_{mirr}$ close to one as shown in the Fig.2 (d). There is always a valley for reflection rate $R^1_{slot}$ of CCW wave in the Fig.2 when $R^1_{mirr}$ varies from zero to one. In a WGM cavity, the asymmetric backscattering of counter-propagating optical waves can be related to chiral behaviour, achieving maximal chirality and unidirectional emission at an exceptional point (EP). Here we call maximal chirality points from Fig.2 as quasi-EPs, which need larger $R^1_{mirr}$ with decrease pump rate r or increase of driving field Rabi frequency $\Omega_\mu$.

After acquiring the value of $R^1_{mirr}$ corresponding to quasi-EPs in the Fig.2, Fig.3 plot slot reflection to counter-propagating waves versus pump rate with driving field Rabi frequency $\Omega_\mu = 1\gamma$ or $\Omega_\mu = 2\gamma$, respectively. With increase of pump rate, reflection rate $R^1_{slot}$ declines dramatically to zero and increases gently to reach a steady value as shown in Fig.3(c) and (d). Otherwise, reflection rate $R^2_{slot}$ rises slowly below one with pump rate. Hence, the chirality behaviour in the WGM microcavity can be adjusted in succession through controlling pump rate to vary dispersion relation.

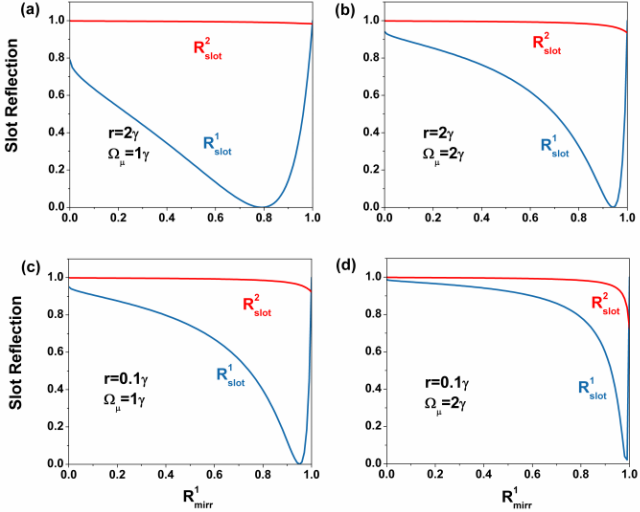

Fig. 2. Slot reflection to CCW ($R^1_{slot}$) and CW ($R^2_{slot}$) waves vary as a function of the reflection of slot surface $R^1_{mirr}$ with different atomic pump rate r and driving field Rabi frequency $\Omega_\mu$, (a) $r=2\gamma$, $\Omega_\mu=1\gamma$ (b) $r=2\gamma$, $\Omega_\mu=2\gamma$ (c) $r=0.1\gamma$, $\Omega_\mu=1\gamma$ (d) $r=0.1\gamma$, $\Omega_\mu=2\gamma$, and the length of the cavity-made slot is set to $L_{slot}=6c\pi/\omega_{12}$ which is 3 wavelengths at $\omega_{12}$.

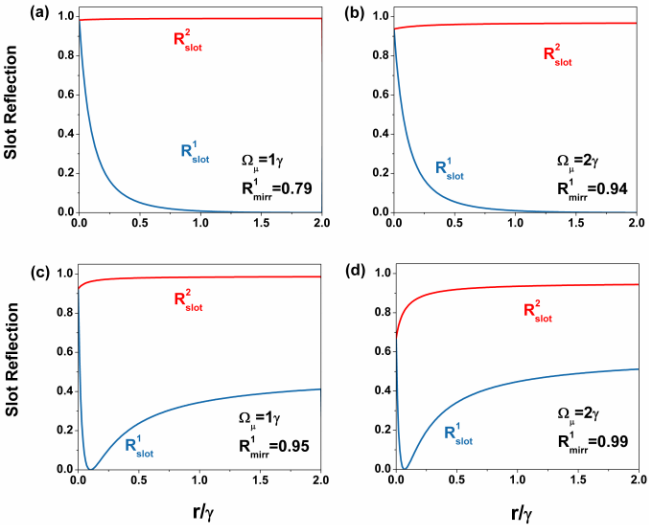

Fig. 3. Slot reflection to CCW ($R^1_{slot}$) and CW ($R^2_{slot}$) waves vary as a function of pump rate r around EP according to Fig.2, (a) $\Omega_\mu=1\gamma$, $R^1_{mirr}=0.79$ (b) $\Omega_\mu=2\gamma$, $R^1_{mirr}=0.94$ (c) $\Omega_\mu=1\gamma$, $R^1_{mirr}=0.95$ (d) $\Omega_\mu=2\gamma$, $R^1_{mirr}=0.99$.

## 3. TUNING CHIRAL MODES

Owing to the cavity-made slot, a portion of the scattered light is lost to the environment, creating an additional damping channel, while the rest couples back into the resonator and leads to the coupling between the counter-propagating WGMs, the degeneracy of which is lifted as a result [18]. This creates two standing wave modes, symmetric mode (SM) and asymmetric mode (ASM) as explained in [19], that are splitted frequency as manifested by the doublet in the transmission spectra in Fig.4, where the SM experiences significant frequency redshift and ASM turns up no shifted frequency. Because the refractive index of cavity-made slot $n_{slot}$ is larger than one, the SM mode always appears on the lower frequency side of the spectrum. The frequency splitting in transmission and reflection spectra is plotted in Fig.4 (a) with pump rate r varying from $0.1\gamma$ to $0.4\gamma$, and in Fig.4 (b) with driving field frequency $\Omega_\mu$ from $7\gamma$ to $10\gamma$. In Fig.4(a) the frequency split is decreased with increase of pump rate r according with that in Fig.5(c), the SM and ASM (see inset of Fig.4) modes are broadened due to the weakness of normal dispersion. Normal dispersion will lead to a narrow resonance linewidth with respect to that of the empty cavity, while anomalous dispersion will lead to a broadening resonance linewidth [17]. In Fig.4(b) the frequency split is increased with increase of driving field frequency $\Omega_\mu$, the SM and ASM modes are narrowed due to the enhancement of normal dispersion.

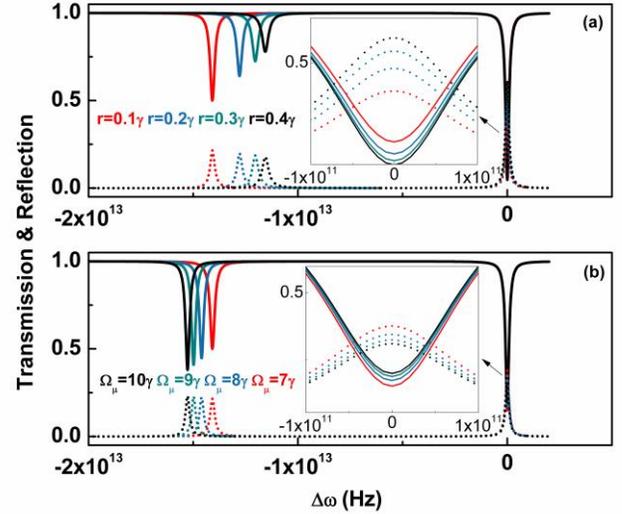

Fig. 4. Transmission and reflection spectra are plotted in (a) with pump rate r varying from $0.1\gamma$ to $0.4\gamma$ and driving field frequency $\Omega_\mu=7\gamma$. (b) Transmission and reflection spectra with driving field frequency $\Omega_\mu$ from $7\gamma$ to $10\gamma$ and $r=0.1\gamma$. Inset of (a) and (b) are the magnified details of no shifted peaks of transmission and reflection spectra.

The eigenmode evolution in WGM microcavity with cavity-made slot broken spatial symmetry exhibits asymmetric backscattering, which is allowed because of the non-Hermiticity of the Hamiltonian [22]. We quantify the imbalance by the chirality in terms of transmission and reflection amplitudes

$$\alpha_{TR} = \frac{||T|-|R||}{|T|+|R|}. \quad (8)$$

In order to ensure the cross-coupling rate $\gamma_c$ in Eq.(3) is not completely decayed in the WGM microcavity, we find the minimum threshold value of driving field frequency $\Omega_\mu^{min}=7\gamma$ with $R^1_{mirr}=0.99$. The chirality $\alpha_{TR}$ of a resonator mode is plotted in Fig.5(b) with different driving field frequency, and asymmetric backscattering of cavity-made slot also are shown in Fig.5(a). The chirality traced back to asymmetric backscattering is continuously tuned via steering the pump rate r of atomic dispersion. The red curve at minimum threshold value $\Omega_\mu^{min}$ has evident elevation with increase of pump rate r and reach smooth after pump rate r is approximately more than $\gamma$, and the maximum chirality is 0.999 close to EP. The tuning range of chiral behavior is shrinked when the driving field is strengthened.

It needs to introduce the frequency splitting quality $Q_{sp}$ to evaluate the experimental observability of the real frequency splitting ref.[20]. According to Eq(4),

$$Q_{sp} = \frac{|\text{Re}(\omega_+) - \text{Re}(\omega_-)|}{-\text{Im}(\omega_+) - \text{Im}(\omega_-)}. \quad (9)$$

In Fig.5(c) the frequency splitting quality $Q_{sp}$ is reduced with increase of pump rate $r$, and the larger driving field frequency is, the smaller adjustable range of $Q_{sp}$ is. In the normal dispersion region shaded in Fig.5, the chirality and $Q_{sp}$ have a prominent variation when the dispersion is tuning by the atomic pump rate. In Fig.5 it also confirms that if chirality is elevated close to one the frequency splitting quality $Q_{sp}$ obtains minimum value, due to both eigenvalues and eigenstates coalesce at a non-Hermitian EP. Moreover, it has been discussed in ref.[20, 23, 24] that in passive systems $Q_{sp}$ <1 the frequency splitting is resolved difficultly in experiments (e.g., in transmission and reflection spectra). Hence it has to introduce gain into systems, such as doping $Er^{3+}$ in the WGM microcavities, to narrow the splitting linewidth for overcoming this experimental difficulty. However, in our dispersive system $Q_{sp}$ >25 as shown in Fig.5 (c), it not only avoids introducing gain into system, but also has a conspicuous frequency split.

Next, we consider the displacement sensitivity of our system in response to the length variation of cavity-made slot. For example, a breathing mechanical breathing mode described as an expansion and contraction of the slot may be excited in our system. The transmission and reflection spectra share similar characteristics. To avoid redundancy, we mainly discuss frequency split in the transmission spectrum induced by change in slot length.

Transmission spectra as a function of the slot length $L_{slot}$ are shown in Fig.6 with different pump rate $r$ in Fig.6 (a) and different diving field frequency $\Omega_\mu$ in Fig.6 (b). The degeneracy of WGM microcavity is lifted due to cavity-made slot, which creates two standing wave modes, symmetric mode (SM) and asymmetric mode (ASM). Both in Fig.6 (a) and (b), the increase of frequency splitting are gradually reduced with the slot length $L_{slot}$. This is because the change in slot length had a major impact on SM, while almost no influence on ASM. When the change of $\Delta L_{slot}$ is increased, the impact on SM deviates the center (antinode) of SM and closes to node. The frequency split is increased with decrease of pump rate $r$ in Fig.6 (a), due to the enhancement of normal dispersion. In Fig.6(b) the frequency split is decreased with decrease of driving field frequency $\Omega_\mu$, due to the weakness of normal dispersion. According to Fig.6, the maximum splitting response is 15.22 THz/nm at $\Omega_\mu$=7 and $r$=0.4$\gamma$.

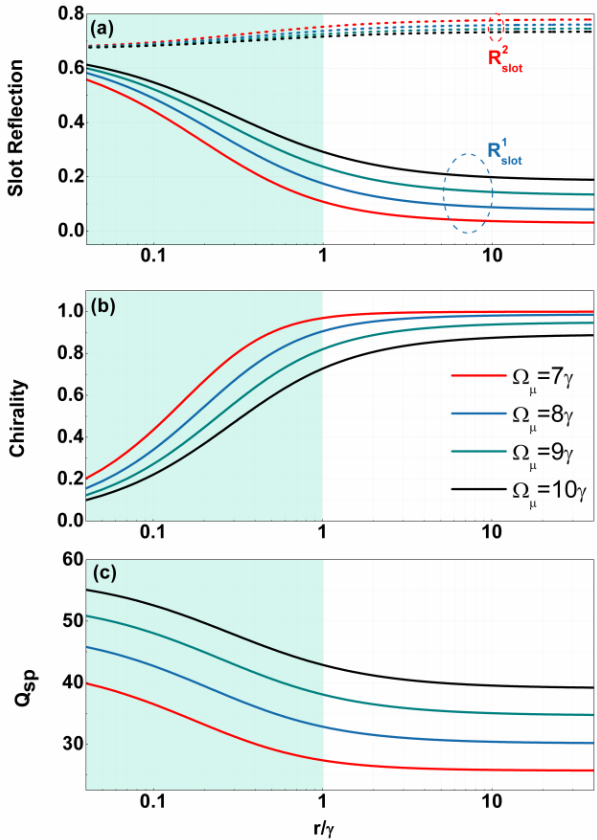

Fig. 5. (b) The chirality is continuously tuned via steering the pump rate $r$. (c) The frequency splitting quality versus pump rate. The shaded regions represent normal dispersion, the other is anomalous dispersion. Some parameters are set by $\kappa_0$=$\kappa_{ex}$=1×10$^{11}$ Hz, $R^1_{mirr}$=0.99, $R^2_{mirr}$=0.999, the radius of WGM microcavity $R_0$=30$\lambda$/(2$\pi$n), refractive index $n$=1.5.

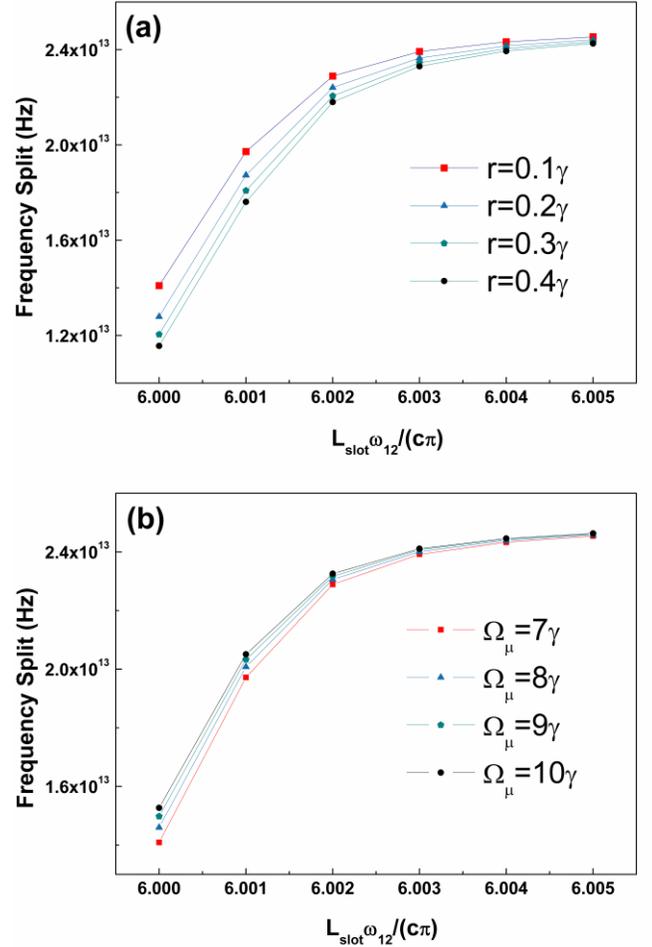

Fig. 6. Transmission frequency split (a) driving field frequency $\Omega_\mu$=7$\gamma$. (b) pump rate $r$=0.1$\gamma$.

## 4. CONCLUSIONS

In summary, a whispering gallery mode resonator with a cavity-made slot filled atomic vapor is demonstrated to present tunable chiral modes. The underlying physical mechanism that exhibits chirality is the asymmetric backscattering in the vicinity of an EP, which occurs in a non-Hermitian system. The chirality behaviour in the WGM microcavity can be adjusted in succession through controlling pump rate to vary dispersion relation. In our system $Q_{sp}$ >25, it does not need to introduce gain into system for observing a conspicuous frequency split. We also analyse the displacement sensitivity of our system in response to the length variation of cavity-made slot. The system exhibits high displacement sensitivity 15.22 THz/nm could be applied to optomechanical sensing applications.

**Funding Information.** This work was supported by the Beijing Academy of Quantum Information Sciences under Grant Y18G28, and the National Natural Science Foundation of China under Grant 11574021, Grant 11574017, Grant 51372008, and Grant 11804017.